# Thermoelectric Hall Effect at High-Magnetic-Field Quantum Limit in Graphite as a Nodal-Line Semimetal

Toshihito Osada*, Tomotaka Ochi, and Toshihiro Taen

*Institute for Solid State Physics, University of Tokyo,*

*5-1-5 Kashiwanoha, Kashiwa, Chiba 277-8581, Japan.*

The high-magnetic-field thermoelectric effect in nodal-line semimetals with straight nodal lines was investigated. Three-dimensional (3D) Dirac/Weyl semimetals exhibit constant thermoelectric Hall conductivity at the high-magnetic-field quantum limit, resulting in a boundless linear increase of Seebeck coefficient. This is known as the quantized thermoelectric Hall effect (QTHE), and is expected to lead to high-performance thermoelectricity at low temperatures. Here, in addition to Dirac/Weyl semimetals, we demonstrated that 3D semimetals with straight nodal lines can also exhibit the QTHE under high magnetic fields. As a candidate material for experimental validation, we discussed the thermoelectric properties of bulk graphite. Furthermore, we investigated the dimensional crossover of thermoelectricity to two-dimensional behavior in thin-film graphite.

The thermoelectricity of topological semimetals at the high-magnetic-field quantum limit has recently attracted considerable attention because of its possible high performance even at low temperatures. It has been shown that, at the clean limit, it has been shown that the thermoelectric Hall conductivity $\alpha_{xy}^{(3D)}$ of three-dimensional (3D) Dirac or Weyl semimetals takes a constant value that is proportional to the temperature, but independent of the magnetic field and/or carrier density. This quantized thermoelectric Hall effect (QTHE) is attributed to the constant density of states of the $N = 0$ chiral Landau sub-bands [1,2]. Correspondingly, the Seebeck coefficient $S_{xx}$ increases linearly with no saturation as the magnetic field increases, resulting in high-performance thermopower. This feature has been observed in 3D Dirac semimetal $Pb_{1-x}Sn_xSe$ with a small spin-orbit gap [3], 3D Dirac semimetal $ZrTe_5$ [4], and Weyl semimetal $TaP$ [5].

Similar thermoelectricity has also been discussed for the two-dimensional (2D) massless Dirac fermion system [6]. Generally, in the 2D electron gas at the quantum Hall region, each Landau level (LL) with a single spin in a single valley contributes $(\log 2)k_B e/h$ to the 2D thermoelectric Hall conductivity $\alpha_{xy}^{(2D)}$, when the chemical potential $\mu$ is located at the LL center [7–9]. Therefore, when the spin and valley degeneracy breaking of the $N = 0$ LL can be ignored, $\alpha_{xy}^{(2D)}$ takes a universal value of $4(\log 2)k_B e/h$ at the high magnetic field limit because $\mu$ is located at the center of $N=0$ LL. This value is independent of the magnetic field, temperature, and carrier density. This QTHE leads to high-performance thermoelectricity, even at low temperatures. However, when the Zeeman splitting breaks the degeneracy of $N = 0$ LL, the QTHE disappears, causing a bump-like structure of the field dependence of $\alpha_{xy}^{(2D)}$ [10].

In this paper, first, we demonstrate that 3D semimetals with straight Dirac nodal lines, which can be realized by stacking 2D Dirac fermion layers, exhibit the QTHE similar to Dirac/Weyl semimetals. This means that high-performance thermoelectricity due to QTHE can be expected in other materials than Dirac/Weyl semimetals. Based on this fact, we indicate that bulk graphite exhibits the QTHE at the clean limit. In addition, we discuss the dimensional crossover of QTHE from 3D to 2D in thin-film graphite.

First, we discuss the thermoelectricity for the following band model:

$$E_\pm(\mathbf{k}) = \pm \hbar v_F \sqrt{k_x^2 + k_y^2} - 2t_c \cos c k_z. \tag{1}$$

Dispersion (1) represents a Dirac nodal-line semimetal with straight nodal lines, where the 2D Dirac fermion layers stack along the $z$-direction with interlayer transfer energy $t_c$. $v_F$ and $c$ are the in-plane Dirac fermion velocity and interlayer spacing (lattice constant), respectively. "$\pm$" represents the conduction (+) and valence (−) bands that linearly touch with each other along the $k_z$ axis to form a straight nodal line. We assume that the system has a two-fold valley degeneracy and two-fold spin degeneracy. A schematic of the dispersion with two valleys is shown in the inset of Fig. 1(b). The Landau sub-bands under a magnetic field along the $z$-axis (normal to 2D layers) are given by

$$E_N(k_z) = \text{sgn}(N)\sqrt{2e\hbar v_F^2 |B_z||N|} - 2t_c \cos c k_z, \tag{2}$$

where $N$ denotes the Landau index ($N = 0, \pm 1, \pm 2, ...$).

The Landau sub-band dispersion (2) at a fixed magnetic field is shown in Fig. 1(a). The sub-bands with $N > 0$ ($N < 0$) correspond to those of the conduction (valence) band, whereas the $N = 0$ sub-band is characteristic of the Dirac system. At the quantum limit, the Fermi level crosses only the $N = 0$ subband. This configuration is essentially the same as that of Dirac/Weyl semimetals at the quantum limit, where the Fermi level crosses only the $N = 0$ chiral Landau sub-bands of the two valleys, as shown in the inset of Fig. 1(a).

We studied the thermoelectricity of this nodal-line system at the clean limit, following previous works on the QTHE in Dirac/Weyl semimetals [1,2]. The thermoelectric Hall conductivity $\alpha_{xy}$ is an off-diagonal element of the thermoelectric conductivity tensor $\vec{\alpha}$ defined by $\mathbf{j} = \vec{\sigma}\mathbf{E} + \vec{\alpha}(-\nabla T)$, where $\mathbf{j}$, $\mathbf{E}$, $\vec{\sigma}$, and $-\nabla T$ are the current density, electric field, electrical conductivity tensor, and temperature gradient, respectively. Note that the dimensions of $\alpha_{xy}$ differ between 2D and 3D systems ($\alpha_{xy}^{(2D)} = \alpha_{xy}^{(3D)} c$). In the dissipationless clean limit, $\alpha_{xy}^{(3D)}$ is expressed as follows [1,2]:

$$\alpha_{xy}^{(3D)} = \frac{1}{c} \cdot \frac{c}{2\pi} \int_{-\pi/c}^{\pi/c} \alpha_{xy}^{\square}(k_z) dk_z , \qquad (3)$$

$$\alpha_{xy}^{\square}(k_z) = 4 \cdot \frac{k_B e}{h} \sum_N \left[ -f^0(E_N(k_z)) \log f^0(E_N(k_z)) - \{1 - f^0(E_N(k_z))\} \log\{1 - f^0(E_N(k_z))\} \right]. \quad (4)$$

In the finite dissipation case, the above formulae provide the leading term of the expansion of $\alpha_{xy}^{(3D)}$ with respect to scattering. Here, factor 4 in (4) corresponds to four-fold spin and valley degeneracy, and $f^0(E) = 1/[1+\exp\{(E-\mu)/k_B T\}]$ is the Fermi distribution function. Eq. (4) represents the 2D thermoelectric Hall conductivity of a virtual 2D system with a fixed parameter $k_z$ [6–10]. The chemical potential $\mu$ included in $f^0(E_N(k_z))$ is determined by the following condition for the imbalance between electron and hole density:

$$n^{(3D)} - p^{(3D)} = 4 \cdot \frac{1}{2\pi l^2} \frac{1}{c} \cdot \frac{c}{2\pi} \int_{-\pi/c}^{\pi/c} \sum_N \{f^0(E_N(k_z)) - \theta(N)\} dk_z . \qquad (5)$$

Here, $l = \sqrt{\hbar/e|B_z|}$ is the magnetic length, and $1/(2\pi l^2)$ is the guiding center degeneracy of the Landau levels. $\theta(N)$ takes the values 1, 1/2, and 0 for $N < 0$, $N = 0$, and $N > 0$, respectively. The value of $\theta(N)$ reflects the filling of the $N$-th Landau sub-band at the high-magnetic-field limit.

Figure 1(b) shows the calculated magnetic-field dependence of $\alpha_{xy}^{(3D)}/T$ at several temperatures at a fixed carrier imbalance $n^{(3D)} - p^{(3D)}$. At low temperatures ($k_B T \ll t_c$), each curve converges to a single curve, except for quantum oscillations, implying that $\alpha_{xy}^{(3D)}$ is almost proportional to $T$. At the quantum limit (yellow shaded region), $\alpha_{xy}^{(3D)}$ takes a constant value

$$\alpha_{xy}^{(3D)} \sim \frac{e k_B^2 T}{3 \hbar c t_c} = \frac{2\pi k_B T}{3 t_c} \frac{k_B e}{ch}, \tag{6}$$

to form a plateau-like structure. This is the same feature as that observed in Dirac/Weyl semimetals. Therefore, the manifestation of QTHE is not limited to Dirac/Weyl semimetals, but also in straight nodal-line systems.

Because the thermoelectric power tensor $\ddot{S}$ is defined by $\mathbf{E} = \ddot{\sigma}^{-1}\mathbf{j} + \ddot{S}\nabla T$, $\ddot{\alpha}$ can be obtained from the experimentally measured $\ddot{\sigma}$ and $\ddot{S}$ using the relation $\ddot{\alpha} = \ddot{\sigma} \cdot \ddot{S}$. If the diagonal elements of $\ddot{\sigma}$ and $\ddot{\alpha}$ are negligibly small at the clean limit, the Seebeck coefficient is approximately written as $S_{xx} \sim \alpha_{xy}^{(3D)}/\sigma_{xy}^{(3D)} = -\alpha_{xy}^{(3D)} B_z / e(n^{(3D)} - p^{(3D)})$. Therefore, as a result of the constant $\alpha_{xy}^{(3D)}$, the boundless linear increase of $S_{xx}$ under high magnetic fields is expected, leading to high-performance thermoelectricity.

Next, we investigated the thermoelectricity of graphite as an example of semimetallic systems with straight nodal lines. Zhu *et al.* experimentally investigated the thermoelectric effect of bulk graphite under magnetic fields [11]. They have found that the quantum oscillation of the Nernst coefficient $S_{xy}$ is significantly enhanced in bulk graphite, in contrast to graphene that exhibits 2D behaviors [12,13]. In contrast, in the present study, we focus on the thermoelectric Hall conductivity $\alpha_{xy}^{(3D)}$ at the clean limit.

To simulate the electronic structure of bulk graphite, we employed the Slonczewski–Weiss–McClure (SWMc) model [14–16] that provides the **k**-expansion of bands around the H-K-H (or H'-K'-H') line in the hexagonal Brillouin zone (Fig. 3(a)). The band dispersion around the H-K-H line is shown in Fig. 3(b). The conduction and valence bands touch along the H-K-H line, forming a straight nodal line with quadratic band contact. Exactly to say, the two bands touch along four Dirac nodal lines almost parallel to H-K-H, if the trigonal warping cannot be ignored [17, 18].

The Landau subband dispersion under the magnetic fields parallel to the H-K-H direction (stacking direction) can be calculated based on the SWMc model [19, 20]. For simplicity, we ignore the trigonal warping by setting one SWMc parameter $\gamma_3$ as zero (cylindrical approximation). As other SWMc parameters, we employed $\gamma_0 = 3.16$ eV, $\gamma_1 = 0.39$ eV, $\gamma_2 = -0.020$ eV, $\gamma_4 = 0.044$ eV, $\gamma_5 = 0.038$ eV, and $\Delta = -0.008$ eV [16]. As lattice constants, $a = 0.246$ nm and $c/2 = 0.337$ nm were used. The carrier imbalance $n^{(3D)} - p^{(3D)} = 1 \times 10^{18}$ cm$^{-3}$ is assumed so as to reproduce quantum oscillations in bulk. The calculated Landau sub-band dispersion at a fixed magnetic field is represented by the solid red curves in Fig. 2(c). Here, $k_z$ is the wave number along the K-H-K direction. The sub-bands labeled $N' = 1e, N' = 2e, \ldots$ originate from the conduction band, whereas those labeled $N' = 1h, N' = 2h, \ldots$ originate from the valence band. The $N' = 0$ and $N' = -1$ sub-bands correspond to the doubly degenerate Landau level at the band contact point in bilayer graphene that is the stacking unit of bulk graphite. The overall sub-band configuration is similar to that of the previous model shown in Fig. 1(a), except the $N'=0$ and $N'=-1$ sub-bands instead of the $N=0$ sub-band in Fig. 1(a).

The thermoelectric Hall conductivity $\alpha_{xy}^{(3D)}$ of bulk graphite was calculated by replacing $N=\{1, 2, ...\}$ with $N'=\{1e, 2e, ...\}$, $N=\{-1, -2, ...\}$ with $N'=\{1h, 2h, ...\}$, and $N=0$ with $N'=\{-1, 0\}$ in (4) and (5). Figure 3(a) shows the magnetic field dependence of $\alpha_{xy}^{(3D)}$ at several temperatures. As the magnetic field increases, $\alpha_{xy}^{(3D)}$ first decreases, exhibiting quantum oscillations, and subsequently shows a QTHE plateau in the quantum limit, as expected. $\alpha_{xy}^{(3D)}$ is almost proportional to the temperature, except for quantum oscillation. Because the width of $N'=-1$ sub-band $4t_c$ is 40 meV in Fig. 2(c), the plateau value can be estimated as $\alpha_{xy}^{(3D)} / (k_B e / ch) \sim 2 \times 2\pi k_B T / 3t_c = 0.036 \, [\text{K}^{-1}] \times T$ using (6). Here, we assumed the same $t_c$ for the $N'=0$ sub-band. The plateau shown in Fig. 3(a) is in good agreement with this estimation.

In addition, we investigated the quantum size effect on the thermoelectric Hall conductivity of thin-film graphite. We considered a thin film consisting of $N_{BL}$ bilayers ($2N_{BL}$ layers). Each Landau sub-band $E_N(k_z)$ in bulk graphite was discretized to the $N_{BL}$ energy levels in thin-film graphite. These levels are represented as $E_N(k_z^{(j)})$, where

$$k_z^{(j)} \equiv \frac{\pi}{c} \frac{j}{N_{BL}+1} \quad (j = 1, 2, ..., N_{BL}) \quad (7)$$

under the fixed-end boundary condition, for a sufficiently large $N_{BL}$. These levels are indicated by solid blue circles in Fig. 2(c). Note that the above assumption fails for ultrathin films, such as few-layer graphene with a small $N_{BL}$.

The 2D thermoelectric Hall conductivity $\alpha_{xy}^{(2D)}$ of thin-film graphite is given by

$$\alpha_{xy}^{(2D)} = \sum_{j=1}^{N_{BL}} \alpha_{xy}^{\square}(k_z^{(j)}), \tag{8}$$

$$n^{(2D)} - p^{(2D)} = 4 \cdot \frac{1}{2\pi l^2} \cdot \sum_{j=1}^{N_{BL}} \sum_{N'} \left\{ f^0(E_{N'}(k_z^{(j)})) - \theta(N') \right\}. \tag{9}$$

Here, $N$ in (4) must be replaced by the Landau index $N'$ of graphite, as mentioned above. Figure 3(b) shows the magnetic field dependence of $\alpha_{xy}^{(2D)}$ of thin-film graphite at several temperatures. At high temperatures ($k_B T \gg t_c / N_{BL}$), $\alpha_{xy}^{(2D)}$ exhibits the same behavior as that of $\alpha_{xy}^{(3D)}$ of bulk graphite ($\alpha_{xy}^{(2D)} / N_{BL} \sim c \cdot \alpha_{xy}^{(3D)}$). Because $\alpha_{xy}^{(2D)}$ is proportional to temperature, the QTHE plateau value decreases with temperature. On the other hand, at low temperatures ($k_B T \ll t_c / N_{BL}$), fine quantum oscillations different from bulk oscillations appear to originate from discretized quantum levels, and the plateau-like structure is no longer sustained in the bulk quantum limit region. According to previous studies on thermoelectricity in a clean 2D system [7–10], each discretized level contributes $(4\log 2)k_B e / h$ to $\alpha_{xy}^{(2D)}$, including spin and valley degeneracy. Therefore, $\alpha_{xy}^{(2D)}$ converges to one oscillatory curve at low temperatures, and each oscillation peak is quantized to $(4\log 2)k_B e / h$ except for overlap region of two oscillations, as shown in Fig. 3(b).

As mentioned above, the thermoelectric Hall conductivity of thin-film graphite shows a crossover from 3D to 2D behavior depending on the temperature and thickness. This dimensional crossover is considered to occur approximately when the QTHE plateau of $\alpha_{xy}^{(2D)}$ reaches the quantized value $(4\log 2)k_B e / h$. This condition is given as follows:

$$2 \times \frac{2\pi k_B T}{3 t_c} < \frac{4\log 2}{N_{BL}}. \tag{10}$$

Using $t_c = 10$ meV, (10) can be written as $T < 76.8$ [K]$/ N_{BL} = 51.8$ [K·nm]$/d$, where $d$ denotes the thickness of the graphite film.

Finally, we discuss an actual experiment using dissipative samples. We have discussed the QTHE of $\alpha_{xy}^{(3D)}$ in clean bulk graphite. At the clean limit, the Seebeck coefficient $S_{xx}$ is expected to increase linearly without saturation under high magnetic fields. This behavior cannot be sustained in dissipative systems because the diagonal electrical and thermoelectric conductivities ($\sigma_{xx}^{(3D)}$ and $\alpha_{xx}^{(3D)}$) become finite because of scattering. However, the QTHE plateau-like behavior of $\alpha_{xy}^{(3D)}$ still survives as a dissipationless part even in the dissipation system. Therefore, $\alpha_{xy}^{(3D)}$ must be evaluated using the measured $\sigma_{xx}^{(3D)}$, $\sigma_{xy}^{(3D)}$, $S_{xx}$, and $S_{xy}$ to experimentally confirm the QTHE [4, 5].

In conclusion, we studied the high-magnetic-field thermoelectric effect in semimetals with straight nodal lines. It is found that $\alpha_{xy}^{(3D)}$ exhibits plateau-like behavior (QTHE) at the quantum limit, similar to the Dirac/Weyl semimetals. This implies that the QTHE and the resulting high-performance thermoelectricity are not limited to the Dirac/Weyl semimetals, but expected in wider class of semimetals. Graphite is one of the nodal-line semimetals where the QTHE is expected.

**Acknowledgements**

The authors thank Dr. A. Kiswandhi, Dr. M. Sato, and Dr. K. Uchida for the valuable discussions. This work was supported by JSPS KAKENHI Grant Numbers JP19K14655, JP20H01860, and JP21K18594.

# References

*osada@issp.u-tokyo.ac.jp

**Figure 1** (Osada et al.)

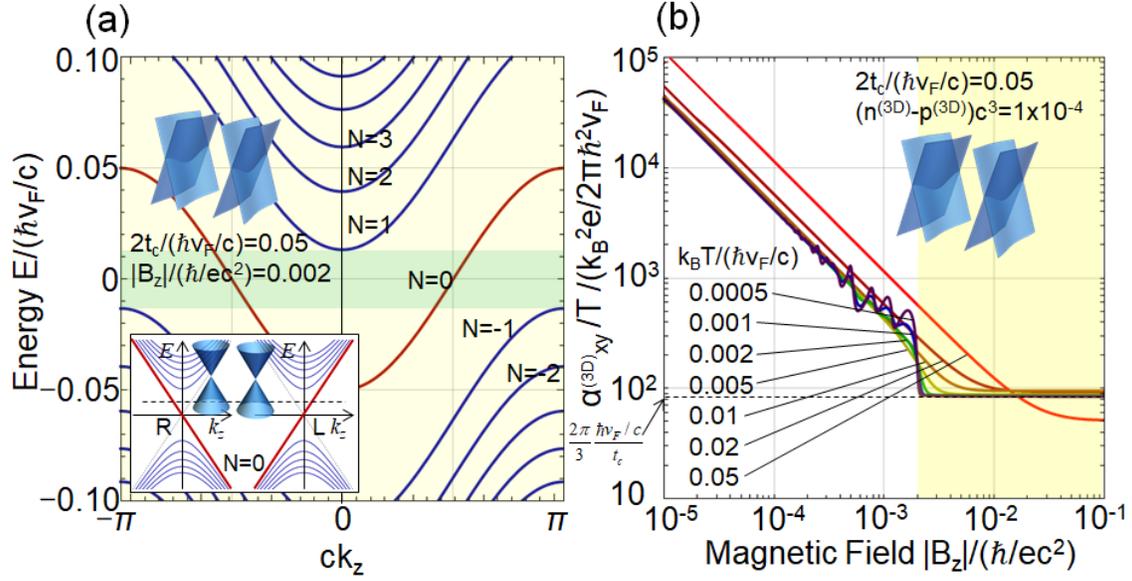

**FIG. 1.** (color online)

(a) Landau sub-band configuration of a Dirac nodal-line semimetal under the magnetic field parallel to the straight nodal lines. Inset: Schematic Landau subband configuration in Dirac/Weyl semimetals. (b) Magnetic field dependence of the thermoelectric Hall conductivity normalized by temperature. Dashed line indicates the plateau value of the QTHE. Inset shows the schematic band dispersion of the system.

**Figure 2** (Osada et al.)

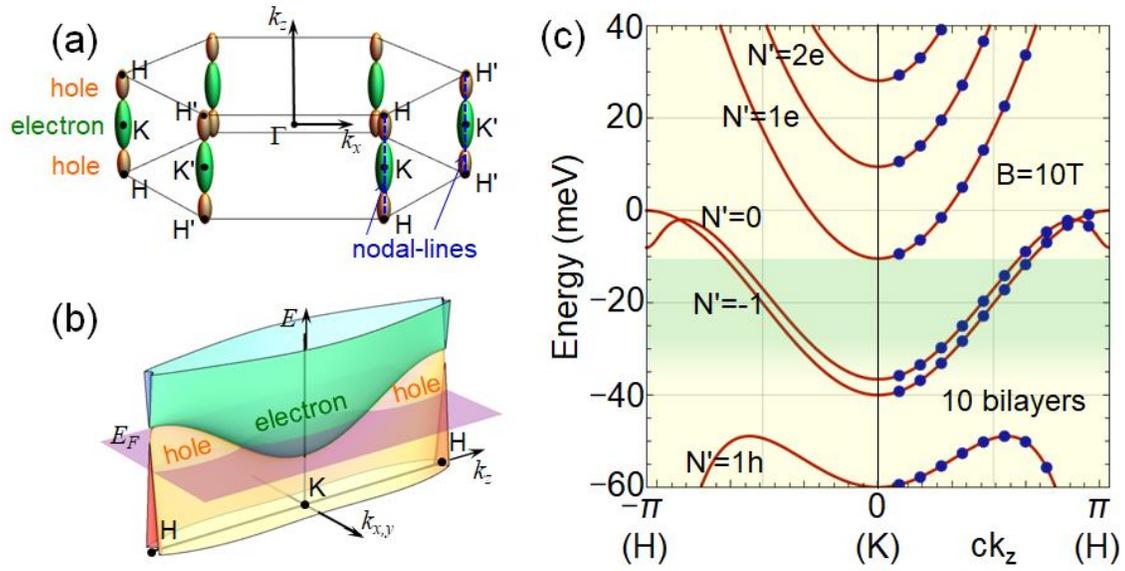

**FIG. 2.** (color online)

(a) Schematic of the Brillouon zone and Fermi surface of graphite. (b) Band dispersion around the H-K-H line in the Brillouin zone. The conduction and valence bands touch quadratically along the H-K-H line. (c) Landau subband configuration of bulk graphite under the magnetic field parallel to the stacking axis ($z$-axis). The solid circles indicate the discretized energy levels in thin-film graphite with $N_{BL} = 10$.

**Figure 3** (Osada et al.)

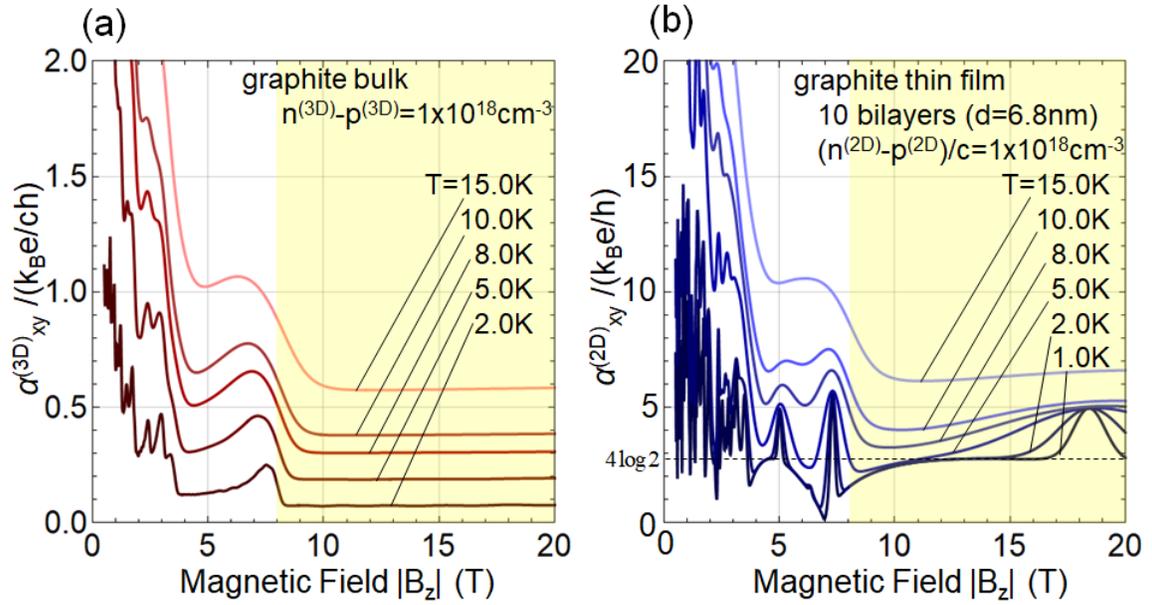

**FIG. 3.** (color online)

(a) Magnetic field dependence of the 3D thermoelectric Hall conductivity $\alpha_{xy}^{(3D)}$ of bulk graphite for several temperatures. (b) Magnetic field dependence of the 2D thermoelectric Hall conductivity $\alpha_{xy}^{(2D)}$ of thin-film graphite with $N_{BL}$ = 10. Dashed line indicates the quantized value of oscillation peak in 2D system.